# Effects of pressure-induced phase transitions on superconductivity in the single crystal $Fe_{1.02}Se$

V. G. Tissen, E. G. Ponyatovskii, M. V. Nefedova

*Institute of Solid State Physics, 142432 Chernogolovka, Moscow Region, Russia*

A. N. Titov, V.V. Fedorenko

*Institute of Metal Physics, RAS Ural Division, 620219 Ekaterinburg, GSP-170, Russia*

**Abstract**

The temperature dependence of the ac susceptibility for single-crystalline $Fe_{1.02}Se$ has been measured at high pressures to 6.7 GPa. The superconducting transition temperature $T_c$ displays a large increase from 7 K to 30 K in the two-phase region at 0.5 – 1 GPa. Further increase in pressure results in the decrease of $T_c$. The destruction of superconductivity near 7 GPa is presumably related to a structural transition from orthorhombic to hexagonal.
.

1. Introduction

The composition of $Fe_{1+x}Se$, the simplest member of the new class of Fe-based high-temperature superconductors, is extremely sensitive to the preparation procedure due to the easy evaporation of Se. Initially, it was thought that superconductivity in this compound exists over a rather wide compositional range [1]. But, as was established later, small deviations in the stoichiometry reduces strongly the temperature of the superconducting transition $T_c$ which falls below 0.6 K at x=0.03 [2].

Despite the close similarity of the basic structural features between $Fe_{1+x}Se$ and the other Fe-based superconductors, the maximal $T_c \sim 8$ K in $Fe_{1+x}Se$ at ambient pressure is much lower than in quaternary compounds containing arsenic where $T_c$ exceeds 30 K. Remarkably, $T_c$ of $Fe_{1+x}Se$ is shifted to this value by the application of high pressures of several GPa [3-7]. Most published results display an interesting feature: the superconducting transition becomes very broad in some range of pressure. Especially, this is clearly seen in the experimental data for single-crystalline samples in the hydrostatic environment [7]. Such a broadening of a resistive superconducting transition was explained by the appearance of the second superconductive state around 2.5 GPa [5].

We show here that the single crystal $Fe_{1.02}Se$ transforms in two distinct superconducting phases coexisting over a narrow range of pressure. Also, the destruction of superconductivity occurs quite abruptly near 7 GPa. Both conclusions are reached from measurements of the ac magnetic susceptibility.

2. Experimental

Single crystals of $Fe_{1+x}Se$ were grown using polycrystalline $Fe_{1+x}Se$ and the iodine vapor transport method. Ceramic $Fe_{1+x}Se$ samples were produced by ampoule synthesis at temperatures below 723 K, as, according to the phase diagram of the Fe-Se system [8], the temperature of decomposition for the phase with tetragonal structure of PbO-type is 730 K. Such a low temperature of synthesis requires a long time to reach phase equilibrium. Therefore, the material after annealing was homogenized by means of intermediate grinding and pressing. We found that after four such cycles the material was predominantly single phase. Because of this long and complex procedure of synthesis, the composition of the material had a deviation from the nominal one mainly due to the evaporation of selenium. Evaporated selenium may subside on the internal ampoule surface. The composition of ceramic samples was controlled by energy-dispersive analysis using JEOL-733 microscope. We should note that this method of analysis, being applied even with etalons, always gives a systematic error related to the overestimation of the content for heavy elements in comparison with lightweight elements. The size of this error was taken into account using $TiSe_2$ single crystals. This compound has no homogeneity range with selenium excess [9], so the composition of most stoichiometric $TiSe_2$ samples was assigned the nominal composition. As the atomic weight of Fe is higher than that of Ti, such a correction also gives a rough estimation only, but it decreases the error in the determination of composition.

The phase composition of $Fe_{1+x}Se$ ceramic was controlled by powder X-ray diffraction using a Shimadzu XRD 7000 diffractometer (Cu $K_\alpha$-radiation, curved monochromator, reflection mode, $2\vartheta$ range 8-90 degrees) performed in "Ural-M" center of the Institute of Metallurgy UrD RAS. According to diffraction data, the main phase of the samples was $Fe_{1+x}Se$ with tetragonal structure of PbO type; the content of impurity phases was less than 2%.

Crystal growth was carried out in quartz ampoules at temperature gradient from 973 K at the hot end to 573 K at the cold end of an ampoule. The length of an ampoule was 200-250 mm. The time of growth was 2-4 weeks. Crystals had typical dimensions from 0.2 mm to 2 mm and different compositions which depend on the position of the crystal in the ampoule and therefore, on the temperature of growth. General tendency is the increase of Se content with decreasing temperature of growth. The crystals grown at the hot end of the ampoule had the content close to $Fe_{1.1}Se$ but were inhomogeneous and contained some amount of metallic iron and $Fe_{1+x}Se$ phase with tetragonal structure, while the main phase of these crystals had a hexagonal structure. Perhaps, this composition is a result of non-equilibrium processes accompanying the decomposition of high-temperature $Fe_{1+x}Se$ phase upon cooling.

The $Fe_{1+x}Se$ of interest to us with tetragonal structure were grown in a narrow ampoule zone, where the temperature was close to 723 K. Their composition was close to stoichiometric FeSe. The structure of the single crystals obtained was studied with a four-circle Oxford Xcalibur diffractometer (Mo $K_\alpha$-radiation, graphite monochromator, position-sensitive CCD-detector) in Postovski Institute of Organic Synthesis UrD RAS.

The small crystal of $Fe_{1.02}Se$ for high-pressure experiments was cleaved from the inner part of an as-grown crystal with trapezoidal shape shown in Fig. 1. High pressures were generated by the diamond anvil cell (DAC) made of nonmagnetic NiCrAl alloy. The sample with dimensions 70x70x20 $\mu m^3$ was loaded inside a hole spark-drilled in a NiMo gasket preindented to 80 $\mu m$. Methanol-ethanol mixture was used as the pressure transmitting medium. Pressure was determined by the ruby fluorescence method at room temperature. The superconducting transition was detected inductively using balanced pick-up coils and lock-in amplifier Stanford Research SR830 [10].

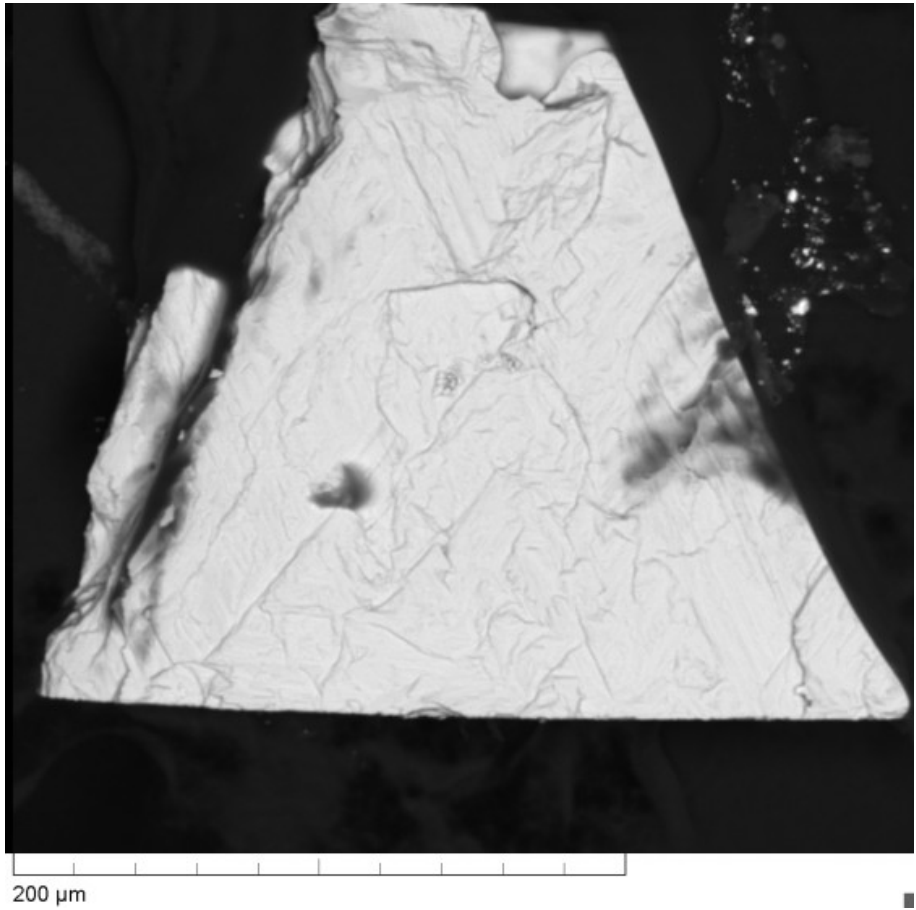

Fig.1. Photograph of the inner part of the crystal of $Fe_{1.02}Se$ which was cut perpendicular to the c axis.

## 3. Results and discussion

The temperature dependences of the ac susceptibility for $Fe_{1.02}Se$ at different pressures are shown in Fig. 2. At ambient pressure, the onset of the superconducting transition at about 7 K is close to that for a polycrystalline sample of the same composition [2]. The application of pressure of 0.2 GPa results in a small decrease of $T_c$. The next increase of pressure to 0.5 GPa changes drastically the shape of the $\chi'(T)$ curves. The amplitude of the signal corresponding to the superconducting transition at about 7 K decreases and the new broad superconducting anomaly appears at higher temperatures. Such a picture is typical for a superconductor undergoing a transition to another phase with higher $T_c$ [11]. As seen in Fig. 3, this $T_c$ rises steeply in the two-phase region. However, this increase in $T_c$ cannot be considered as an intrinsic property of the second superconducting phase because of the essential influence of the proximity effect on $T_c$.

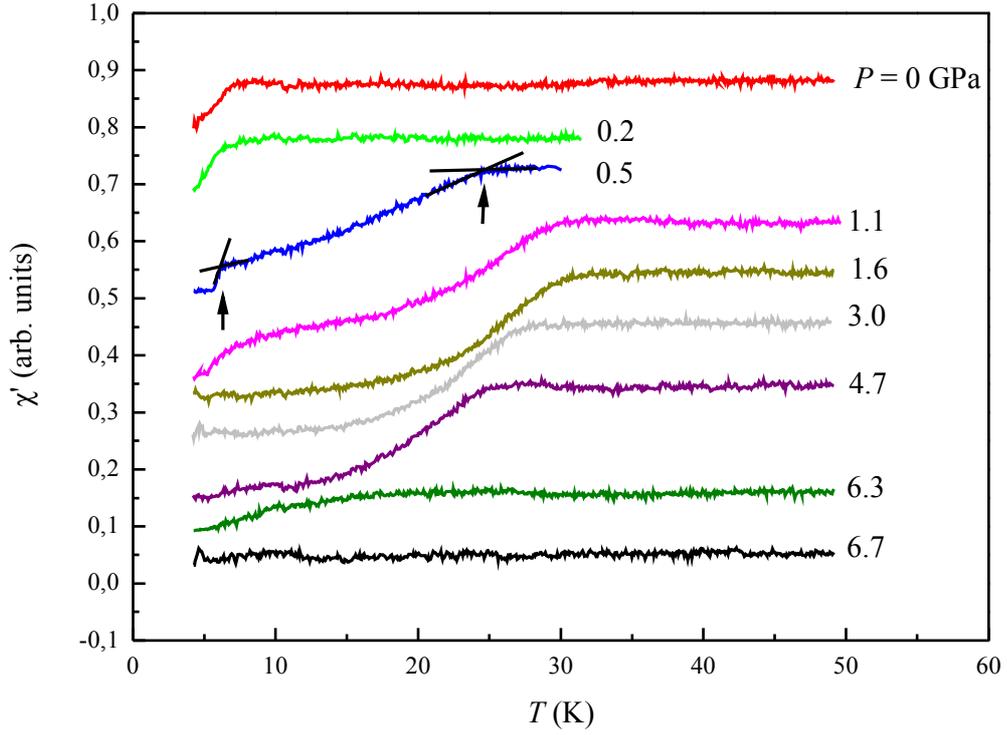

Fig. 2. The ac susceptibility versus temperature for $Fe_{1.02}Se$ at different pressures. Arrows mark the $T_c$ values.

At 1.6 GPa, the sample becomes single-phased with $T_c$ = 30 K, decreasing at higher pressures. Finally, above 4.7 GPa the superconducting signal broadens considerably and shifts downward rapidly, indicating a possible phase transformation to non-superconducting state. At this point, the force was increased to attain an expected pressure of 7.5 GPa. However, the measured value was 6.7 GPa only. This might be ascribed to a substantial drop in the volume of the sample during pressurization. As seen in Fig. 2, no traces of superconductivity are observed at 6.7 GPa in the studied temperature range above 4.2 K.

It is well-established that the location of the maximum of $T_c$ on the pressure scale in iron arsenide superconductors strongly depends on doping level (for a review, see [12, 13]). It seems that the same picture is valid for $Fe_{1+x}Se$. Also, it was found recently that the structural phase transitions at ambient pressure are extremely sensitive to the stoichiometry of $Fe_{1+x}Se$ [14]. The superconducting $Fe_{1.01}Se$ possessing the tetragonal structure at room temperature transforms to the orthorhombic phase at 90 K while the non-superconducting $Fe_{1.03}Se$ keeps the tetragonal structure upon cooling down. Therefore, the value of pressure for the orthorhombic-to-hexagonal structural transition accompanied by a substantial drop in $T_c$ above 9 GPa [15] might vary noticeably for a small change of the concentration in $Fe_{1+x}Se$. The suppression of superconductivity in $Fe_{1.02}Se$ observed in the present study is presumably connected to this transformation occurring at lower pressure due to the difference in composition.

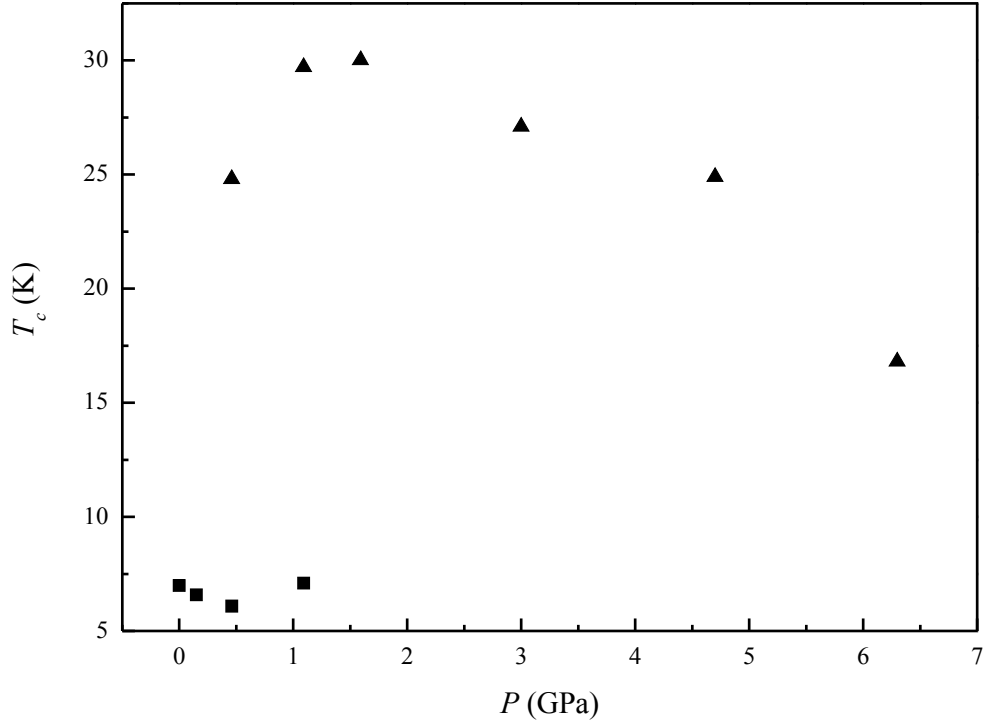

Fig. 3. Superconducting transition temperature of $Fe_{1.02}Se$ versus pressure.

An unusually high compressibility for the orthorhombic phase of $Fe_{1+x}Se$ was revealed under high pressure using synchrotron X-ray diffraction at 16 K [15]. No anomalies are seen in the pressure dependence of the unit cell volume to ~ 14 GPa. However, the height of the Se atoms above the Fe plane drops in a step-like manner at about 2 GPa. If this anomaly were a signature of an isostructural transition, then the two-phase region found in the present study around 1 GPa might be due to a similar transformation.

## 4. Conclusions

The application of pressure enhances $T_c$ of $Fe_{1.02}Se$ from ~ 7 K to ~ 30 K, presumably due to the formation of a new superconducting phase. Structural studies on single crystals $Fe_{1+x}Se$ at low temperatures and high pressures are needed to clarify what changes of structure result in such a large variation of $T_c$.

**Acknowledgements**

The authors wish to thank J. S. Schilling for very valuable comments and corrections which improved the manuscript and E. Z. Kurmaev for informational support. The work at ISSP was supported by the RAS Program (Project 09-02-00709) and the RFBR (Grant 09-02-00709-a). The work at IMP was supported by the Quantum physics of Condensed Matter (Grant No. 4 of the UrD RAS) and the RFBR (Grant 09-03-00053-a).

**References**

[1] F. C. Hsu, J. Y. Luo, K. W. The, T. K. Chen, T. W. Huang, P. M. Wu, Y. C. Lee, Y. L. Huang, Y. Y. Chu, D. C. Yan, and M. K. Wu, Proceedings of the National Academy of Sciences **105**, 14262 (2008).

[2] T. M. McQueen, Q. Huang, V. Ksenofontov, C. Felser, Q. Xu, H. Zandbergen, Y. S. Hor, J. Allred, A. J. Williams, D. Qu, J. Checkelsky, N. P. Ong, and R. J. Cava, Phys. Rev. B **79**, 14522 (2009).

[3] Y. Mizuguchi, F. Tomioka, S. Tsuda, T. Yamaguchi, and Y. Takano, Appl. Phys. Letters **93**, 152505 (2008).

[4] S. Medvedev, T. M. McQueen, I. Trojan, T. Palasyuk, M. I. Eremets, R. J. Cava, S. Naghavi1, F. Casper, V. Ksenofontov, G. Wortmann and C. Felser, arXiv:0903.2143.

[5] V.A. Sidorov, A.V. Tsvyashchenko, R.A. Sadykov, arXiv:0903.2873.

[6] G. Garbarino, A. Sow, P. Lejay, A. Sulpice, P. Toulemonde, W. Crichton, M. Mezouar, and M. Nunez-Regueiro, arXiv:0903.3888.

[7] D. Braithwaite, B. Salce, G. Lapertot, F. Bourdarot, C. Marin, D. Aoki and M. Hanfland, J. Phys.: Condens. Matter **21**, 232202 (2009).

[8] H.Okamoto, J. Phase Equilibrium and Diffusion **12,** 383 (2001).

[9] J.A.Wilson, Phys. stat. sol. (b) **86**, 11 (1978).

[10] I. O. Bashkin, M. V. Nefedova, V. G. Tissen, and E. G. Ponyatovsky, JETP Letters, **80**, 655 (2004).

[11] I. O. Bashkin, V. K. Fedotov, M. V. Nefedova, V. G. Tissen, E. G. Ponyatovsky, A. Schiwek, and W. B. Holzapfel, Phys. Rev. B **68**, 054401 (2003).

[12] C. W. Chu and B. Lorenz, arXiv:0902.0809.

[13] Yu. A. Izyumov and E. Z. Kurmaev, Physics-Uspekhi **51**, 12 (2008).

[14] T. M. McQueen, A. J. Williams, P. W. Stephens, J. Tao, Y. Zhu, V. Ksenofontov, F. Casper, C. Felser, and R. J. Cava, arXiv:0905.1065.

[15] S. Margadonna, Y. Takabayashi, Y. Ohishi, Y. Mizuguchi, Y. Takano, T. Kagayama, T. Nakagawa, M. Takata, and K. Prassides, arXiv:0903.2204.